\documentclass[a4paper,final]{appolb}
 \usepackage[]{graphicx} 
\begin{document}

\pagestyle{plain}
\title{Transverse hydrodynamics in relativistic heavy-ion collisions.
  \thanks{Talk presented at the 2009 Cracow Epiphany Conference}
}
\author{Rados{\l }aw Ryblewski
  \address{The H. Niewodnicza\'nski Institute of Nuclear Physics,\\
    Polish Academy of Sciences,\\
    ul. Radzikowskiego 152, PL-31342 Krak\'ow, Poland}
}
\maketitle
\begin{abstract}
General features of the formalism describing hydrodynamic evolution of transversally thermalized matter possibly produced at the very early stages of ultra-relativistic heavy-ion collisions are presented. Thermodynamical consistency of the model is emphasized. The covariant formulas for the moments of the phase-space distribution function are derived. The simple model for the transition from purely transverse to standard perfect-fluid hydrodynamics is proposed.   
\end{abstract}
\section{Introduction}
\label{sect:Intro}

It is commonly accepted that the evolution of the hot and dense matter created in heavy-ion collisions at the RHIC energies can be reasonably well described in the framework of the relativistic hydrodynamics of a perfect fluid \cite{Teaney:2001av,Huovinen:2001cy,Kolb:2002ve,Hirano:2002ds,Hirano:2004rs,Heinz:2005zg,Nonaka:2006yn,Hama:2005dz,Shuryak:2004cy,Bass:2000ib}. Nevertheless, the standard hydrodynamic approach, assuming three dimensional (3D) thermalization of the system, encounters severe problems concerning initial conditions. The main issue in this respect is to explain the very early thermalization time of the created system, which is required by the standard hydrodynamic models to describe the elliptic flow coefficient and the particle transverse-momentum spectra.

Recently a possible solution of this puzzle has been proposed by Bialas et al. in \cite{Bialas:2007gn}. In this approach one assumes that at the very early stages the hydrodynamic description applies only to transverse degrees of freedom, where the thermalization is quite easy to achieve, while in the longitudinal direction the motion is reduced to simple parton free streeming. The thermalized, two-dimensional (2D) objects corresponding to the group of particles moving with the same rapidity are called (transverse) clusters. 

Originally, the idea of the purely transverse thermalization of matter created in heavy-ion collisions was introduced by Heinz and Wong \cite{Heinz:2002rs,Heinz:2002xf}, with the conclusion that such a model cannot describe the RHIC data. However, the new formulation of this concept in Refs. \cite{Bialas:2007gn,Chojnacki:2007fi} showed that the idea of the purely transverse thermalization may be  consistent with the experimental results.     

In this paper we concentrate on the discussion of the formal aspects of the transverse-hydrodynamics model, which were first studied in \cite{Ryblewski:2008fx}. In particular, we extent formal results obtained in \cite{Bialas:2007gn,Chojnacki:2007fi} by including the effects of finite parton masses and quantum statistics.

\section{Hydrodynamical equations}
\label{sect:HE}

The relativistic hydrodynamic equations of the perfect fluid valid for transversally thermalized matter follow from the energy-momentum conservation law, $\partial_\mu T^{\mu \nu}=0$, with the energy-momentum tensor of the form
\begin{equation}
T^{\mu \nu} = \frac{n_0}{\tau} \left[
\left(\varepsilon _2 + P_2\right) U^{\mu}U^{\nu} 
- P_2 \,\,\left( g^{\mu\nu} + V^{\mu}V^{\nu} \right)\,\, \right].
\label{tensorT1}
\end{equation}
Here $\tau = \sqrt{t^2 - z^2}$ is the longitudinal proper time and $n_0$ describes the density of clusters in rapidity. 

Since clusters are 2D objects, whose thermodynamic properties should be described by the proper 2D thermodynamic variables, we introduce the quantities $\varepsilon_2$, $P_2$, $n_2$ and $s_2$ which denote the 2D energy density, pressure, particle density, and entropy density, respectively. The essential feature of our treatment is that those variables satisfy the standard thermodynamic identities. In our considerations the number of particles is not conserved, hence, throughout the paper we restrict ourselves to the case of vanishing chemical potential, $\mu = 0$. In this case we have
\begin{equation}
\varepsilon_2 + P_2 = T s_2
\label{thermid1}
\end{equation}
and
\begin{equation}
d\varepsilon_2 = T ds_2, \quad dP_2 = s_2 dT,
\label{thermid2}
\end{equation}
where $T$ stands for the temperature.

The definition of the energy-momentum tensor (\ref{tensorT1}) includes also the two mutually orthogonal four-vectors $U^\mu$ and $V^\mu$. The timelike four-vector $U^\mu$, defined by the equation 
\begin{equation}
U^{\mu} = ( u_0 \cosh\eta,u_x,u_y, u_0 \sinh\eta),
\label{U}
\end{equation}
describes the four-velocity of the fluid element and may be obtained from the four-vector $u^\mu = \left(u^0, {\vec u}_\perp, 0 \right) = u^0 \left(1, v_x, v_y, 0 \right) $  by performing the Lorentz boost along the $z$ axis with the spacetime rapidity
\begin{eqnarray}
\eta = \frac{1}{2} \ln \frac{t+z}{t-z}. \label{yandeta}
\end{eqnarray}
The quantity $u^\mu$ is the normalized to unity four-velocity of the fluid element in the rest frame of the cluster and  $u^0 \equiv \gamma = \left(1-v^2\right)^{-\frac{1}{2}}$ is the Lorentz gamma factor, determined by the hydrodynamic transverse flow \mbox{$v = (v_x^2+v_y^2)^{1/2}$}.

The special role played by the longitudinal direction in the transverse hydrodynamics, as the direction of the free streaming of the clusters, entails the appearance of the additional spacelike four-vector $V^\mu$, which is defined by the equation
\begin{equation}
V^{\mu} = (\sinh\eta,0,0,\cosh\eta).
\label{V}
\end{equation}
In the rest frame of the cluster, where $\eta=0$, one finds \mbox{$V^\mu = (0,0,0,1)$}.  The straightforward calculation of the components of the energy-momentum tensor in the rest frame of the fluid element (where in addition to the condition $\eta=0$ we have also ${\vec u}_\perp=0$) yields the diagonal form
\begin{equation}
T^{\mu \nu} = \frac{n_0}{\tau} \left(
\begin{array}{cccc}
\varepsilon _2 & 0 & 0 & 0 \\
0 & P_2 & 0 & 0 \\
0 & 0 & P_2 & 0 \\
0 & 0 & 0 & 0
\end{array} \right).
\end{equation}
Vanishing of the $T^{33}$ component describing the longitudinal pressure is a direct consequence of the appearance of the term $V^\mu V^\nu$ in (\ref{tensorT1}) and it indicates that there is no momentum transport across the cluster, which also means that there is no interaction between the clusters. Furthermore, as compared to the standard hydrodynamics, the lack of the longitudinal pressure effects in a more rapid transverse expansion of the matter, since all the energy accumulated in the clusters at the very beginning of the collision has to be used for transverse acceleration only. Of course, it is expected that such a situation cannot last long since the collisions between partons tend to thermalize the system also in the longitudinal direction. Hence, within at most few fermis after the collision, the transition from transverse to standard 3D hydrodynamics is expected.   

One may check by the direct calculation that the energy-momentum conservation law leads to the entropy conservation law
\begin{equation}
U_\nu \partial_\mu T^{\mu \nu} = T \partial_\mu S^\mu =0,
\label{entcl}
\end{equation}
whose structure is similar to that known from the standard hydrodynamics. In Eq. (\ref{entcl}) we introduced the entropy current which is given by the expression
\begin{equation}
S^\mu = \frac{n_0}{\tau} s_2 U^\mu.
\label{hydroS} 
\end{equation}
Equation (\ref{entcl}) agrees with our expectations, since the evolution in the cluster is entropy conserving and the longitudinal expansion of clusters is just free streaming, which cannot produce entropy. In other words, the assumption that the considered system is a simple superposition of the 2D expanding perfect fluids implies directly the adiabaticity of the flow.

Similarly to the standard hydrodynamics we can incorporate Eq.(\ref{entcl}) into the energy-momentum conservation law to derive the relativistic analog of the Euler equation 
\begin{eqnarray}
U^\mu \partial _\mu (T U^\nu) &=& \partial ^\nu T  + V^\nu V^\mu \partial_\mu T. 
\label{hydro2D}
\end{eqnarray}
In comparison with the standard hydrodynamics, an additional term proportional to $V^\mu V^\nu$ is present in Eq.~(\ref{hydro2D}). By performing projections of Eq.~(\ref{hydro2D}) on the four-vectors $U^\mu$ and $V^\nu$ we check that Eq. (\ref{hydro2D}) contains only two independent equations. Altogether with Eq. (\ref{entcl}) we have a system of three independent equations for four unknown functions 
\begin{eqnarray}
s_2 \left(\tau, {\vec x}_\perp  \right), \quad 
T \left(\tau, {\vec x}_\perp  \right),   \quad 
u_x \left(\tau, {\vec x}_\perp  \right), \quad 
u_y \left(\tau, {\vec x}_\perp  \right).
\label{hydrofunct}
\end{eqnarray}
Hence, as in the standard hydrodynamics, we need to add the equation of state connecting $s_2$ and $T$ to solve the system of hydrodynamic equations. The reduction of the number of independent equations from three to two is expected, because the evolution along the $z$ axis is fixed and only the transverse evolution needs to be determined.

Our hydrodynamic equations can be further rewritten in a different form which is more convenient for numerical calculations. In this case we use standard cylindrical coordinates, i.e., we introduce the distance from the beam axis \mbox{$r\!\!=\!\!\sqrt{r_x^2+r_y^2}$} and the azimuthal angle \mbox{$\phi=\hbox{tan}^{-1} (r_y/r_x)$},  where \mbox{${\vec x}_\perp = (r_x,r_y)$}. We also introduce the parameterization of the fluid velocity in the form
\begin{eqnarray}
v_x &=& v \cos(\alpha+\phi), \quad v_y = v \sin(\alpha+\phi). 
\label{vxvy}
\end{eqnarray}
The dynamically changing angle $\alpha$ describes deviations of the flow from the radial direction. Thus, it specifies the asymmetry of the flow present in the most common non-central collisions. 

\begin{figure*}[t!]
\begin{center}
\includegraphics[width=0.8\textwidth]{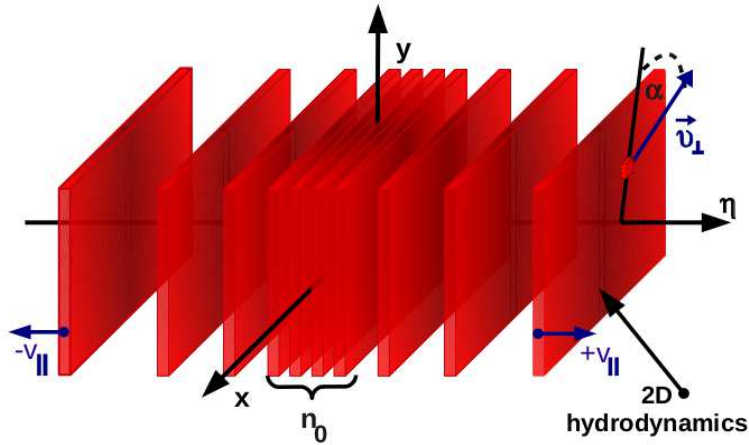}
\end{center}
\caption{Rapidity dependence of the cluster density $n_0$. }
\label{fig:fig1}
\end{figure*}

With the help of the variables introduced above, and using thermodynamic identities only, we obtain the explicit form of the hydrodynamic equations
\begin{eqnarray}
&& \frac{\partial }{\partial \tau} \left( r s_2 u_0 \right) +\frac{\partial }{\partial r}
\left( r s_2 u_0 v\cos \alpha \right) + \frac{\partial }{\partial
\phi }\left( s_2 u_0 v\sin \alpha \right)  =0, \nonumber \\
&& \frac{\partial }{\partial \tau}\left( rTu_0 v\right) +
r\cos \alpha \frac{\partial }{\partial r}\left( Tu_0 \right) 
+\sin \alpha \frac{\partial }{\partial \phi }\left( Tu_0 \right)  =0, \nonumber \\
&& Tu_0 ^{2}v\left( \frac{d\alpha }{d \tau}+\frac{v\sin \alpha }{r}\right) -\sin
\alpha \frac{\partial T}{\partial r}
+\frac{\cos \alpha }{r}\frac{\partial T}{\partial \phi } =0. 
\label{wfdyr3}
\end{eqnarray}
The quantity $d/d\tau$ is the total time derivative
\begin{eqnarray}
\frac{d}{d\tau}=\frac{\partial}{\partial \tau} + v \cos \alpha \frac{\partial}{\partial r} + \frac{v \sin \alpha }{r} \frac{\partial}{\partial \phi} .
\label{totaltd}
\end{eqnarray}  
It turns out that the parameter $n_0$, treated as a constant so far, may be an arbitrary function of the  spacetime rapidity $\eta$. We can show this feature by multiplying the original energy-momentum tensor $T^{\mu \nu}_{n_0 = 1}$ by $n_0=n_0 \left( \eta \right)$ and incorporating it into the energy-momentum conservation laws
\begin{equation}
\partial_\mu \left[ n_0 \left( \eta \right) T^{\mu \nu}_{n_0 = 1} \right] = 0.
\label{tensoreta}
\end{equation}
Since the tensor $T^{\mu \nu}_{n_0 = 1}$ is conserved, Eq.(\ref{tensoreta}) is reduced to the equation
\begin{equation}
T^{\mu \nu}_{n_0 = 1} \, \partial_\mu n_0 \left( \eta \right) = 0,
\label{tensoreta2}
\end{equation}
which, as the straightforward calculation shows, is always fufilled. This feature indicates that our model does not have to be necessarily boost invariant, see Fig. \ref{fig:fig1}. In fact, a closer analysis shows that different initial conditions can be applied to different clusters (placed at different values of $\eta$). This is so because the clusters do not interact and evolve independently of each other.      

Although the full 3D thermalization of the system at the very beginning of the evolution is hard to achieve, as the system expands the thermalization in the longitudinal direction seems to be inevitable. Such situation requires the transition from the transverse hydrodynamics to standard hydrodynamics. Probably, the simplest description of such a transition is achieved with the assumption that it has a sudden character. In this case the transition may be realized by assuming the so called Landau matching conditions at the transition point 
\begin{equation}
T^{\mu \nu} U_\nu = T_{3D}^{\mu \nu} U_\nu.
\label{LMC}
\end{equation}
In Eq. (\ref{LMC}) we used the energy-momentum tensor of the perfect fluid valid for standard hydrodynamics
\begin{equation}
T_{3D}^{\mu \nu} = \left(\varepsilon + P\right) U^\mu U^\nu - P g^{\mu \nu}.
\label{standardTmunu}
\end{equation}
In Eq. (\ref{standardTmunu}) the energy density $\varepsilon$ and pressure $P$ are 3D thermodynamic quantities. Eq. (\ref{LMC}), with the appropriate tensors used, may be reduced to the local conservation of the energy and momentum (at the transition point)
\begin{equation}
\frac{n_0}{\tau} \varepsilon_2 U^\mu  = \varepsilon U^\mu.
\label{LMC1}
\end{equation}
It must be supplemented by the assumption of the entropy production associated with 2D $\to$ 3D transition,
\begin{equation}
\frac{n_0}{\tau} s_2 < s,
\end{equation}
where we introduced the 3D entropy density $s$.

We note that the 2D $\to$ 3D transition has been recently addressed in the framework of the dissipative hydrodynamics in \cite{Bozek:2007di,Bozek:2007qt}. In this framework the transition is naturally associated with the entropy production. On the other hand, the 2D $\to$ 3D transition was also studied in Refs. \cite{Florkowski:2008ag,Florkowski:2009sw} in the framework of the entropy-conserving anisotropic hydrodynamics and magnetohydrodynamics, respectively. Moreover, the Landau matching conditions similar to those advocated above were also used recently to model a transition from the early 3D parton free streaming to 3D perfect-fluid hydrodynamics in Ref. \cite{Broniowski:2008qk}.

\section{Thermodynamics of two-dimenssional systems}
\label{sect:TH}

Thermodynamic variables used in the transverse hydrodynamics are obtained directly from the potential $\Omega$ defined for 2D systems of non-interacting bosons (upper signs) or fermions (lower signs)
\begin{equation}
\Omega(T,V_2,\mu) = \pm \nu _g T V_2 \int \frac{d^2 p_{\perp} }{(2\pi)^2}\,
\ln \left(1 \mp e^{(\mu-m_\perp)/T} \right).
\label{Omega1}
\end{equation}
The particles move in the transverse plane whose area is $V_2$, and their energy is \mbox{$\varepsilon_p = m_\perp = \sqrt{m^2 + p_x^2 + p_y^2}$}. The factor $\nu_g$ denotes internal degrees of freedom. In the case of gluons $\nu_g = 16$. Using Eq. (\ref{Omega1}) we can define particle density $n_2 = N_2/V_2$, pressure $P_2$,  entropy density $s_2 = S_2/V_2$, and energy density $ \varepsilon_2  = E_2/V_2$ by using the standard thermodynamic definitions:
\begin{eqnarray}
N_2 &=& - \left( \frac{ \partial \Omega}{ \partial \mu} \right) _{V_2 , T}, 
\label{N2} \\
P_2 &=& - \left( \frac{ \partial \Omega}{ \partial V_2} \right) _{T, \mu} = - \frac{\Omega}{V_2},    
\label{P2} \\
S_2 &=& - \left( \frac{ \partial \Omega}{ \partial T} \right) _{\mu, V_2},
\label{S2} 
\end{eqnarray}
and the relation
\begin{equation}
E_2 + P_2 V_2 = T S_2 + \mu \, N_2.
\label{GD1}
\end{equation}

Simple analytic formulas may be obtained for massless fermions and bosons 
\begin{eqnarray}
n_{2} &=& \frac{\nu_g \pi T^2}{24},  \quad 
\varepsilon_{2} = \frac{3 \nu_g  \zeta(3) T^3}{4 \pi} \quad \hbox{(fermions)},
\label{fermions} 
\end{eqnarray}
\begin{eqnarray}
n_{2} &=& \frac{\nu_g \pi T^2}{12},  \quad 
\varepsilon_{2} = \frac{ \nu_g  \zeta(3) T^3}{\pi} \quad \hbox{(bosons)}.
\label{bosons} 
\end{eqnarray}
Here $\zeta$ is the Riemann zeta function. For the two statistics we also find that sound velocity $c_s$ is constant and much higher than in the standard 3D hydrodynamics, namely
\begin{eqnarray}
P_2 = \frac{1}{2} \varepsilon_2,  \quad 
c_s^2 = \frac{\partial P_2}{\partial \varepsilon_2} = \frac{1}{2}. \label{pressure} 
\end{eqnarray}
In the case of the Boltzmann statistics, the analytic expressions can be derived also for finite masses
\begin{eqnarray}
n_2 &=& \frac{ \nu _g T}{2\pi} (m+T) e ^{ -m/T },  \label{en2} \\
P_2 &=& \frac{ \nu _g T^2}{2\pi} (m+T) e ^{ -m/T }, \label{pe2} \\
s_2 &=& \frac{ \nu _g}{2\pi} [m^2+3mT+3T^2] e ^{ -m/T}, \label{es2} \\
\varepsilon_2 &=& \frac{ \nu _g T }{2\pi} [T^2+(m+T)^2] e ^{ -m/T }.  \label{eps2}
\end{eqnarray}
For the classical statistics and massive particles the sound velocity is given by the formula
\begin{eqnarray}
c_s^2 = \frac{T(m^2+3 m T + 3 T^2)}{m^3 + 3 m^2 T + 6 m T^2 + 6 T^3}.
\label{cs2-1}
\end{eqnarray}
In the limit $m \to 0$  Eqs. (\ref{en2}) -- (\ref{cs2-1}) are reduced to those given first in Ref.~\cite{Bialas:2007gn}. 

\section{Phase-space distribution function and its moments}
\label{sect:PSDF}

Our formalism is based on the covariant form of the phase-space distribution function that is factorized into the longitudinal and transverse part, 
\begin{equation}
F  = f_{\parallel} \, g.
\label{Fxp}
\end{equation}
The structure of the transverse part, describing the hydrodynamic evolution of a single cluster, comes from the assumption of the local equilibrium
\begin{equation}
g\left(p^\mu U_\mu\right) = \frac{1}{e^{p^\mu U_\mu/T} \mp 1},
\label{geqh}
\end{equation}
where 
\begin{equation}
p^\mu U_\mu = m_\perp u_0 \cosh(y-\eta) - {\vec p}_\perp \cdot {\vec u}_\perp.
\label{pdotU}
\end{equation}
The longitudinal part follows from the assumption about the free streaming of particles 
\begin{eqnarray}
f_{\parallel} =  n_0 \frac{ \delta ( y - \eta )}{m_{\perp } \tau} =
\frac{n_0}{\tau } \delta \left( p^{\mu} V_{\mu} \right),
\label{fpar-1}
\end{eqnarray}
and realizes the condition $y=\eta$ by using the Dirac delta function. By joining together Eqs. (\ref{geqh}) and (\ref{fpar-1}) we obtain the covariant formula
\begin{equation}
F  = \frac{n_0}{\tau } \delta \left( p \cdot V \right) g\left( p \cdot U \right).
\label{Fxp1}
\end{equation}
Having the explicit form of the phase-space distribution function (\ref{Fxp1}) at our disposal we can calculate its moments, i.e., the particle current $N^{\mu}$, the energy-momentum tensor $T^{\mu\nu}$, and the entropy current $S^{\mu}$, defined by the momentum integrals
\begin{eqnarray}
N^{\mu} &=& \frac{n_0 \nu _g }{ (2\pi)^2 \tau}   \int \frac{d^3p}{p^0} p^{\mu} 
\delta (p \cdot V ) g(p \cdot U),		\label{MN}			\\
T^{\mu\nu} &=& \frac{n_0 \nu _g }{ (2\pi)^2 \tau}   \int \frac{d^3p}{p^0} p^{\mu} p^{\nu}
\delta (p \cdot V ) g(p \cdot U),				\label{MT}	\\
S^{\mu} &=& -\frac{n_0 \nu _g}{(2\pi)^2 \tau }	  
 \int \frac{d^3p}{p^0} p^{\mu}\,\,\delta (p \cdot V)  
g(p \cdot U) \left[\,\, \ln[ g ( p \cdot U) ] - 1\,\, \right]. 
\label{MS}
\end{eqnarray}
The structure of Eqs. (\ref{MN}) -- (\ref{MS}) follows from the Lorentz structure of the distribution function. Having in mind its scalar character, we find the following decompositions in terms of the  four-vectors $U^{\mu}$ and $V^{\mu}$, 
\begin{eqnarray}
N^{\mu} &=& a \, V^{\mu} + b \, U^{\mu}, \label{MND} \\
 T^{\mu \nu} &=& a^{\, \prime} U^{\mu}U^{\nu} + b^{\, \prime} g^{\mu \nu} 
+ c^{\, \prime} V^{\mu} V^{\nu} + \frac{d^{\, \prime}}{2} (U^{\mu}V^{\nu} + U^{\nu}V^{\mu}),
\label{MTD} \\
S^{\mu} &=& a^{\, \prime \prime} V^{\mu} + b^{\, \prime \prime} U^{\mu},
\label{MSD}
\end{eqnarray}
where in addition the metric tensor $g^{\mu \nu} = \hbox{diag} \left( +1, -1, -1, -1 \right)$ has been used. Since all the coefficients in Eqs. (\ref{MND}) -- (\ref{MSD}) are scalar quantities, they can be calculated in the rest frame of the fluid by performing the proper projections of Eqs. (\ref{MND}) -- (\ref{MSD}) on the four-vectors  $U^{\mu}$ and $V^{\mu}$. One can check that after simple algebra we obtain Eqs. (\ref{tensorT1}) and (\ref{hydroS}).

\section{Summary}

In this paper we have discussed the general formalism of transverse hydrodynamics. The thermodynamic consistency of the model has been demonstrated in the straightforward calculations using only thermodynamic identities. The hydrodynamical equations have been derived and written in the form useful for numerical calculations. The explicit form of the moments of the phase-space distribution function has been obtained with the method based on the covariant tensor decomposition. The transition to the three-dimensional hydrodynamics has been proposed that uses the Landau matching conditions. With the appropriate implementation of the transition in the numerical code we hope to successfully describe the physical observables measured at RHIC.  

\end{document}